\documentclass[epl,twocolumn,amssymb,amsmath,nobibnotes]{revtex4}

\usepackage{graphicx}
\usepackage{color}
\usepackage{subfigure}
\usepackage{dcolumn}
\usepackage{bm}
\usepackage{threeparttable}
\usepackage{stmaryrd}
\begin{document}

\newenvironment{figurehere}
  {\def\@captype{figure}}
  {}

\title{Effect of interfacial strain on spin injection and spin polarization of Co$_{2}$CrAl/NaNbO$_{3}$/Co$_{2}$CrAl magnetic tunneling junction}
\author{Yongqing Cai$^1$}
\author{Zhaoqiang Bai$^{1,2}$}
\author{Ming Yang$^1$}
\author{Yuan Ping Feng$^1$}
\email{phyfyp@nus.edu.sg}
\affiliation{
	$^1$Department of Physics, National University of Singapore,
		2 Science Drive 3, Singapore 117542, Singapore\\
	$^2$Data Storage Institute, Agency for Science, Technology and Research, 5 Engineering Drive 1, Singapore 117608, Singapore}
\date{\today}

\begin{abstract}
First-principles calculations were carried out to investigate interfacial strain effects on spin injection and spin polarization of a magnetic tunnel junction consisting of half-metallic full-Heusler alloy Co$_{2}$CrAl and ferroelectric perovskite NaNbO$_{3}$. Spin-dependent coherent tunneling was calculated within the framework of non-equilibrium Green's function technique. Both spin polarization and tunnel magnetoresistance (TMR) are affected by the interfacial strain but their responses to compressive and tensile strains are different. Spin polarization across the interface is fully preserved under a compressive strain due to stronger coupling between interfacial atoms, whereas a tensile strain significantly enhances interface states and lead to substantial drops in spin polarization and TMR. 
\end{abstract}

\maketitle
\section{Introduction}
Injection of spin-polarized electrons or holes from one material to another has been the subject of intensive studies because it is essential in many spintronic device applications such as magnetic tunneling junction (MTJ), spin valves and spin injectors\cite{SP_GuoH,Felser_Review,BaiZQ_APL}. In general, there are two major approaches to achieve a high spin injection efficiency. One is to use a semiconducting or insulating barrier to selectively filter carriers with one spin state from normal ferromagnetic metals such as Co or Fe, as demonstrated in the Fe/MgO/Fe MTJ\cite{Fe_MgO}. The other is to adopt a half metal as the spin source\cite{Miura_PRB,Chadov_PRL,GapOrigin}. For the latter approach, since there exists a gap for one spin state, the spin polarization (SP) at the Fermi energy can reach 100\% and a high spin injection efficiency can be achieved without the need of a spin filter. Besides the high spin polarization, another significant advantage of using half metals instead of ordinary metallic ferromagnets as a spin source is the elimination of impedance mismatch between the spin source and the semiconductor which is essential for efficient spin injection\cite{Impedance_Mismatch}.

Among the various half metals proposed for spin source, the full-Heusler alloys (A$_{2}$BC) in the L2$_{1}$ structure have attracted great interest due to their large magnetic moments per unit cell\cite{Chadov_PRL,GapOrigin}. In this type of alloys, there exists a band gap for minority spin arising from hybridization between the 3$d$ orbitals of A and B atoms and interaction between the $d$ orbitals of the nearest neighbor
type A atoms\cite{GapOrigin}. Extensive theoretical and experimental studies have been carried out to explore high SP in several full-Heusler alloys\cite{Chadov_PRL,MixMode_MgO,GapOrigin,Impedance_Mismatch}. For example, a spin-injection efficiency of around 50\% was achieved
from a full-Heusler alloy, Co$_{2}$FeSi, to (Al,Ga)As and MgO\cite{Co2FeSi_diodes,Co2FeSi_SP}. A relatively large tunnel magnetoresistance (TMR) has been reported for Co-based full-Heusler alloys such as Co$_{2}$Cr$_{1-x}$Fe$_{x}$Al\cite{GapOrigin,Jourdan_APL}.

The measured SP of bulk Heusler alloys, however, is usually much lower than theoretical predictions. Most of bulk Heusler alloys lose their high SP due to phase disorder and defects\cite{Defects}. More importantly, when a Heusler alloy is heterogeneously grown on another material, interface states arising from charge-discontinuity at the interface dramatically reduces its SP and TMR ratio\cite{InterfaceStates_2011}.
In addition, SP could be affected by interfacial strain due to lattice mismatch between the Heusler alloy and the semiconductor/insulator. Theoretical calculations have shown that strain can alter the width of the minority band gap and the position of the Fermi level in the minority band gap\cite{CCA_Strain}. Botters \emph{et al.}\cite{Strain_MagAnistropy1} demonstrated that a small strain (0.08\%) applied to half-Heusler alloy NiMnSb induces a large change (20\%) in its magnetic anisotropy. A similar strain-induced effect was also observed in Co$_{2}$MnGa\cite{Strain_MagAnistropy2}. On the other hand, it is much harder to control the interfacial strain experimentally due to complex film growth/deposition processes. Relatively little is known about interfacial strain effect besides it
varies depending on the thickness of the film\cite{thickness}.
The performance of a spintronic device based on heterostructures is dominated by interface quality. While the quality of an interface can be due to many factors, interfacial strain is critical and often the cause of other interface imperfections such as interfacial defects. It is therefore critically important to systematically investigate and quantitatively understand the effect of interfacial strain on SP and TMR of Heusler alloy/semiconductor junctions\cite{LatticeStrain,Lee_Nature,Pressure_sensor}.

In the present work, we evaluate effect of an interfacial strain on spin injection of a Heusler alloy-based junction, Co$_{2}$CrAl/NaNbO$_{3}$/Co$_{2}$CrAl (CCA/NNO/CCA), using a first-principles computational approach. We choose Co$_{2}$CrAl because of its remarkable magnetoresistive property and persistence of its half-metallic behavior at  surface and interface\cite{GapOrigin,Galanakis_CCA}. On the other hand, sodium niobate is a suitable spacer layer because of its broad ferroelectric and piezoelectric applications and good lattice match with Co$_{2}$CrAl. We demonstrate that both SP and TMR are affected by the strain but their responses to positive (tensile) and negative (compressive) strains are different.

\section{Calculation Details}
First-principles calculations based on spin-polarized density functional theory were carried out using the VASP package\cite{vasp}. The projector augmented wave method was used to describe electron-ion interaction while the Perdew-Burke-Ernzerhof generalized gradient approximations (PAW-PBE) was adopted for electron exchange and correlation. A cutoff energy of 400 eV and a 6$\times6\times$1 Monkhorst-Pack \textbf{k}-mesh were chosen which provide a good balance between accuracy and computational time. Structural relaxations were performed until the Hellmann-Feynman force on each atom became less than 0.02 eV/{\AA}. The CCA(001)/NNO(001) heterostructure is formed by directly stacking CCA(001) on NNO(001) without rotation, as shown in Fig.~1. There is a small lattice mismatch of 2.4\% between CCA and NNO. We assume that the Heusler alloy is grown on the top of the semiconductor\cite{Co2FeSi_diodes}, and therefore, reduced slightly the in-plane lattice constant of cubic CCA to match those of NNO, as it is often done in interface modeling\cite{Impedance_Mismatch}.
The lattice parameter of the supercell in the direction normal to the interface was adjusted until the total energy of the system reached a minimum, which resulted in a $c/a$ of 1.05 for CCA. Odd numbers of CCA (19) and NNO (9) atomic planes were used in the model to prevent formation of artificial electrostatic fields across the interfaces.

\begin{figure}[b]
\centering
\includegraphics[width=8.5cm]{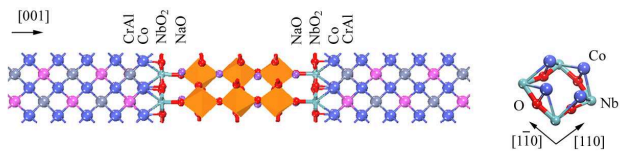}\\
\caption{(Color online) Atomic structure of the CCA/NNO/CCA MTJ. The models are built by stacking CCA directly on NNO along the [001] crystalline direction.}
\label{FIG.1}
\end{figure}

\section{Results and discussion}
Since CCA consists of alternate Co and CrAl planes in the [001] direction, and similarly NNO is composed of alternate NbO$_{2}$ and NaO planes, a CCA(001)/NNO(001) interface can be built with Co- or CrAl-terminated CCA and NbO$_{2}$- or NaO-terminated NNO. In addition to the four interface models (models 1-4 in Table I) built with various combinations of CCA(001) and NNO(001) terminations, we consider another model (model 5) which is formed by CrAl-terminated CCA and NaO-terminated NNO, but the CCA is shifted related to NNO in the interface plane so that Cr forms bond with O
and Al is bonded to Na, instead of the Al-O and Cr-Na bonds in the interface built by directly stacking CCA on NNO (model 4). To evaluate the relative stability of the various interface models, we calculate the interfacial cohesive energy ($E\rm_{ice}$) which is defined as
\begin{eqnarray}
W\rm_{ice} &=& (E\rm_{CCA}+E\rm_{NNO}-E\rm_{CCA/NNO/CCA})/2A
\end{eqnarray}
where $E\rm_{CCA}$, $E\rm_{NNO}$, and $E\rm_{CCA/NNO/CCA}$ are total energies of the CCA slab, the NNO slab, and the CCA/NNO/CCA supercell, respectively. $2A$ accounts for the total area of the 2 interfaces in the supercell. The calculated interfacial cohesive energies, along with the lengths of atomic bonds at the interfaces, are listed in Table~I. The positive and significant values of the interfacial cohesive energies indicate that all five interfaces are stable which is consistent with the fact that  Heusler alloys can form stabile junctions with distinct interfacial structures. It is interesting to note that both CrAl-NaO interfaces are surprisingly more stable than the CrAl-NbO$_{2}$ interface which has double the number of interfacial oxygen bonds. The strongest CCA/NNO interface is the Co-NbO$_{2}$ interface with O atoms lying on top of Co atoms as shown in Fig.~1. The large cohesive energy of this interface (2.46 J/m$^{2}$) is accompanied by a strong Co-O bond. The Co-O bond length (1.94 {\AA}) is close to that in bulk CoO. Since the binding energy of this interface is 20\% larger than that of the next most stable interface, this model is used in subsequent calculations of electronic structure, SP and spin injection.

\begin{table}[htbp]
\caption{\label{tab:results} Calculated interfacial cohesive energies ($W\rm_{ice}$) and interface bond lengths ($R$) for the five CCA/NNO interfaces being studied.}
\begin{ruledtabular}
\begin{tabular}{lcll}
 Interface           &        $W\rm_{ice}$ (J/m$^2$)     &      \multicolumn{2}{c}{$R$    (\AA)}          \\  %
\hline
 1: Co-NbO$_{2}$        &           2.46             &           Co-O (1.94)&   Co-Nb (2.71)           \\
 2: Co-NaO              &           1.72             &           Co-O (1.85)&   Co-Na (2.68)           \\
 3: CrAl-NbO$_{2}$      &           1.64             &           Al-O (1.99)&   Cr-O  (1.97)           \\
 4: CrAl-NaO            &           2.01             &           Al-O (1.83)&   Cr-Na (2.81)           \\
 5: CrAl-NaO            &           1.93             &           Cr-O (1.89)&   Al-Na (2.91)           \\
\end{tabular}
\end{ruledtabular}
\end{table}

The spin-polarized local density of states (LDOS) of the CCA/NNO/CCA MTJ is shown in Fig.~2. The strain introduced by matching the lateral lattice constant of CCA to those of NNO induces minor changes in the magnetic property of CCA, compared to that of bulk cubic CCA. The calculated local magnetic moment of Co is 0.73 $\mu_{B}$ and that of orthorhombic Cr is 1.61 $\mu_{B}$, which are very close to those in cubic CCA (0.77 $\mu_{B}$ for Co and 1.54 $\mu_{B}$ for Cr), and the total magnetic moment is in good agreement with prediction (3 $\mu_{B}$) based on the Slater-Pauling scheme, $M_{t}=Z_{t}-24$, where $M_{t}$ is the total magnetic moment (in $\mu_{B}$) and $Z_{t}$ is the total number of valence electrons in the unit cell.

\begin{figure}[t]
  \centering
  \includegraphics[width=7.5cm]{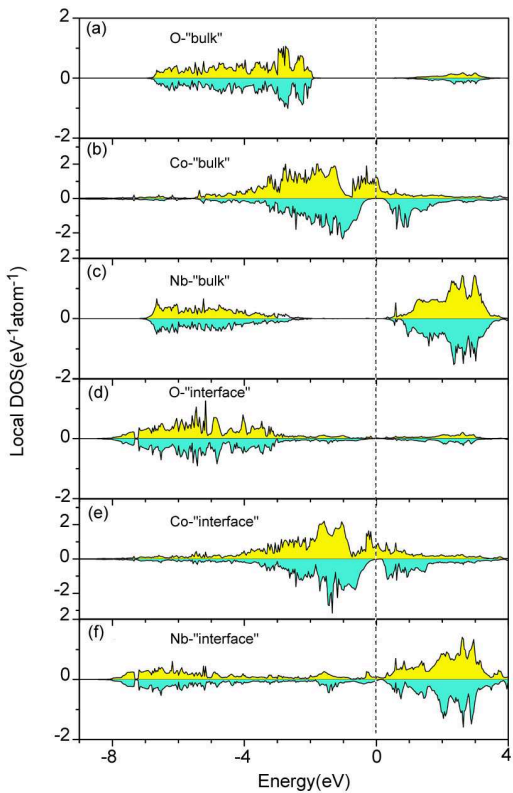}\\
  \caption{(Color online) LDOS for the majority (top panels) and minority (bottom panels)
spin electrons projected to the O, Co and Nb atoms at the Co-NbO$_{2}$ interface and
in bulk region, respectively. The vertical dashed line indicates the position of the
Fermi level.}\label{FIG.2}
\end{figure}

The most important feature of the LDOS is that the half-metallic property of CCA is preserved at the interface, despite of substantial modifications of the electronic structures of the interfacial atoms compared to those in the bulk. For instance, the magnetic moments of the interfacial Co (0.39 $\mu_{B}$) and Cr (1.11 $\mu_{B}$) become smaller than the corresponding values in bulk CCA. The minority band gap is reduced from $\sim$0.5 eV (GGA value) in bulk CCA to $\sim$0.2 eV due to orbital mixing between the Co $3d$ and O $2p$ orbitals. This strong hybridization pushes down the bonding state of the interfacial O atoms by nearly 1 eV (Fig.~2d). In addition to these bonding states between O and Co atoms in the low energy part of the $3d$ states of the interfacial Co atoms, the oxygen LDOS displays a broad band of antibonding states, ranging from $-3$ eV to 3 eV, which are absent in bulk NNO. Similar metal induced gap states (MIGS) appear more prominently for the interfacial Nb atoms as shown in Fig.~2f. Since the length of the Nb-Co bond is 2.71 {\AA} and the Wigner radii of Nb and Co atoms are 1.27 and 1.30 {\AA} respectively, direct bonding between the interfacial Nb and Co atoms seems unlikely, and these MIGS on Nb atoms must arise from an indirect coupling between Nb and Co mediated by O atom. This indirect coupling leads to a substantial exchange-splitting of the states in the energy range from the Fermi energy down to about 2 eV below (Fig.~2f), and a magnetic moment of 0.05 $\mu_{B}$ on the interfacial Nb atom which is aligned antiparallel to those of the Co atoms.
This situation is quite similar to the case of Co/SrTiO$_{3}$ junction where the interfacial Ti atom is found to possess a small magnetic moment which is antiferromagnetically coupled to that of the interfacial Co atom\cite{CoSTO}.

\begin{figure}[h]
  \centering
 \includegraphics[width=6.0cm]{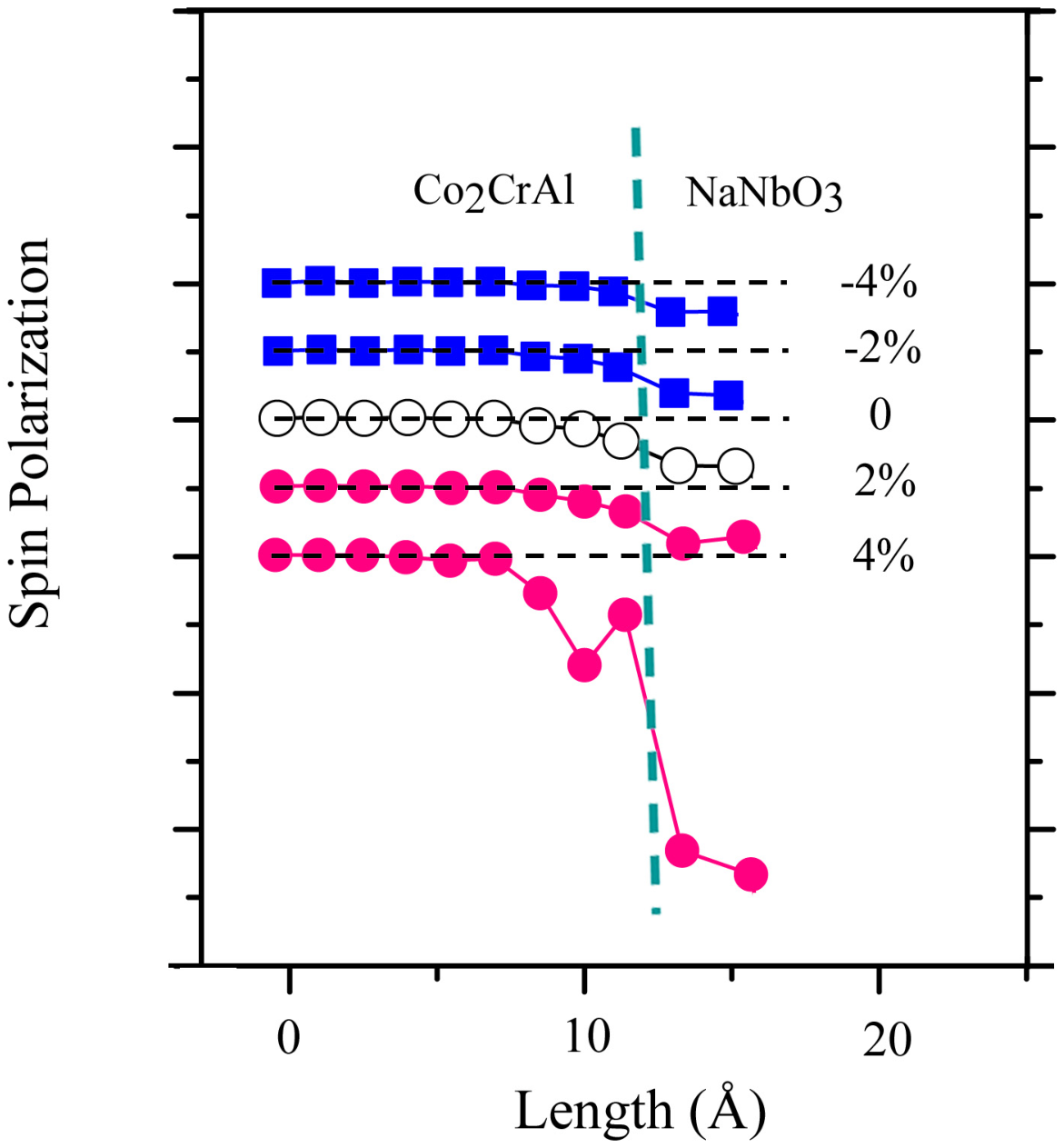}\\
  \caption{(Color online) Variation of the planar averaged SP along the interface
normal direction of the CCA/NNO/CCA MTJ, at various strains.
The position of the Co-NbO$_2$ interface plane is indicated by the vertical
dashed line. The horizontal dashed line corresponds to 100\% SP in each case.}\label{FIG.3}
\end{figure}

Next we discuss the strain effects on the SP and electronic tunneling of the MTJ. A uniaxial strain along the normal direction of the interface is considered and is realized in our model through changing the length of the supercell while fixing the atomic positions in the extreme left and extreme right unit cells of CCA in the supercell (Fig.~1). Contraction or expansion in the lateral directions are estimated to be small in the range of unaxial strain in the present study and are neglected. The system is then relaxed under the constraint of fixed supercell parameter $c$. The SP of each atom at the Fermi level is defined as
\begin{eqnarray}
SP&=& \frac{\rho_{\uparrow}-\rho_{\downarrow}}{\rho_{\uparrow}+\rho_{\downarrow}}
\end{eqnarray}
where $\rho_{\uparrow}$ and $\rho_{\downarrow}$ are the majority and minority LDOS, respectively, at the Fermi level, projected on the concerned atoms. The calculated in-plane averaged SP is shown in Fig.~3 for various strain values. Since the structure and SP are nearly symmetric with respect to the middle of the MTJ structure, only  SP for half of the MTJ is shown in the figure.
A 100\% SP can be observed for atoms at the CCA side, reflecting the half-metallic character of CCA. At zero strain, an extremely high SP (nearly 100\%) is seen at the Co-NbO$_{2}$ interface which extends to  NNO. Under a compressive strain, a steady increase of SP in the NNO part can be clearly observed. This is due to
stronger coupling between interfacial atoms under a compressive strain. Since the minority band gap of CCA originates from the $e_{u}$-$t_{1u}$ splitting due to interaction between the nearest neighbor Co atoms\cite{GapOrigin}, the splitting is enhanced under a compressive strain due to increased hybridization. Under a moderate positive strain, SP in  NNO  drops slightly while the polarization
in CCA remains high ($>$90\%). However, when the strain approaches $\sim$4\%, the SP decreases dramatically. This cannot be attributed to structural instabilities since no significant structural changes occur in the heterostructure at this strain and the Co-O bond length increases only slightly, from 1.94 {\AA} at zero strain to 1.95 {\AA} at
4\% strain.

\begin{figure}[h]
  \centering
   \includegraphics[width=8.0cm]{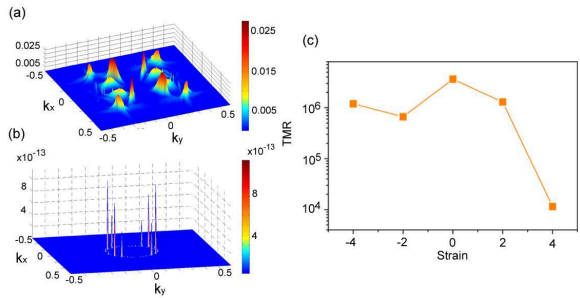}\\
  \caption{(Color online) In-plane wave vector $k_{\sslash}=(k_{x},k_{y})$ dependence
of majority spin (a) and minority spin (b) transmissions at the Fermi level of the
CCA/NNO/CCA MTJ in the parallel magnetization configuration without strain.
(c) TMR ratio as a function of strain.}\label{FIG.4}
\end{figure}

According to Julli\`{e}re\cite{Julliere}, SP is a key quantity in determining TMR of a MTJ. The asymmetrical response of SP to positive and negative strains in the MTJ prompted us to calculate the electron tunneling and TMR across the junction\cite{Conductance_Details}. The TMR ratio is defined as
\[
{\rm TMR}=\frac{G_{\uparrow\uparrow}-G_{\uparrow\downarrow}}{G_{\uparrow\downarrow}}
\]
where G$_{\uparrow\uparrow}$ and G$_{\uparrow\downarrow}$ are conductances of the MTJ when the magnetizations of the two electrodes are parallel and antiparallel, respectively. The conductance per unit cell area is given by
the Landauer-B\"{u}ttiker formula
\[
G=\frac{e^{2}}{h}\sum_{\sigma k_{\sslash}}T_{\sigma}(k_{\sslash})
\]
where $T_{\sigma}(k_{\sslash})$ is the transmission probability of the electron with spin $\sigma$ at the Fermi energy, with $k_{\sslash}=(k_{x},k_{y})$ being the Bloch wave vector in the plane of the junction. The calculated $k_{\sslash}$-resolved transmission in the parallel magnetization configuration is shown in Fig.~4. Similar to the case of the CCA/MgO system\cite{CCA_MgO}, no transmission was found for the  majority spin states around the $\Gamma$ point.
For the minority spin states, the transmission shows spiky structures, but of much smaller magnitudes, which can be attributed to the resonant
tunneling through the weak interface states. The strong transmission of the majority states and much smaller and spiky transmission peaks for the minority states suggest that a high TMR ratio can be expected. The calculated TMR ratio shown in Fig.~4c confirms this prediction. Furthermore, the TMR ratio fluctuates with compressive strain, but decreases rapidly with the tensile strain. At 4\% of tensile strain, the TMR ratio drops by two orders of magnitude compared to the strain free value. This result is in good agreement with recent experimental observation\cite{Pressure_Heusler} that the TMR of a Co$_{2}$MnSi-based MTJ fluctuates slightly, without any abrupt changes, in the process of applying a hydrostatic pressure on the system. To date, no experiment has been conducted to investigate the effect of a positive strain on such a system.

\begin{figure}[h]
  \centering
   \includegraphics[width=8.0cm]{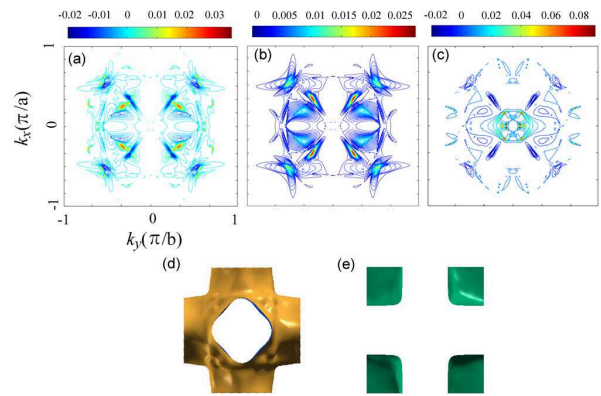}\\
  \caption{(Color online) The differences between majority spin transmissions at the
Fermi level in the parallel magnetization configuration in strained and strain free
CCA/NNO/CCA MTJ: (a) $G_{\uparrow\uparrow}(-4\%)- G_{\uparrow\uparrow}(0)$,
(c) $G_{\uparrow\uparrow}(4\%)- G_{\uparrow\uparrow}(0)$. The transmission at zero
strain is given in (b). The Fermi surfaces corresponding to the two folded bands
in the Brillouin zone of the orthorhombic Co$_{2}$CrAl are shown in (d) and (e),
respectively.}\label{FIG.5}
\end{figure}

The features of the transmission spectrum are related to the shape of the Fermi surface folded into the first Brillouin zone of the deformed orthorhombic CCA supercell.
The difference between the $k_{\sslash}$-resolved majority transmissions of the strained MTJ and the strain free MTJ in the parallel magnetic configuration is shown in Fig.~5a for $-4$\% strain and in Fig.~5c for 4\% strain, respectively. At a strain of $-4$\%, large changes are seen at the highly conducting $k$-points compared to transmission at
zero strain which is shown in Fig.~5b. These $k$-dependent changes can be qualitatively understood by variations of tunneling probability of electrons at the Fermi surface. In contrast, at 4\% strain, new transmission channels, which are absent in the transmission spectrum of the strain free MTJ, are found in the central region of the Brillouin zone where a hole appears in the Fermi surface (Figs.~5d and 5e). These additional conducting channels are related to the interface states resulting from hybridization between atomic orbitals of interface atoms due to the strain induced
interface structural changes. The interface states also greatly enhance the conductance of the minority electrons and destroy the half-metallic character of CCA at the interface. Since the conductance in the antiparallel magnetic configuration of the electrodes G$_{\uparrow\downarrow}$($k_{\sslash}$) can be approximated by the geometrical
mean of the majority and minority conductances obtained for the parallel configuration, {\em i.e.}, $G_{\uparrow\downarrow}(k_{\sslash})=\sqrt{G_{\uparrow\uparrow}(k_{\sslash}) G_{\downarrow\downarrow}(k_{\sslash})}$, one can expect $G_{\uparrow\downarrow}$($k_{\sslash}$) to increase dramatically for states around $\Gamma$ and a substantial reduction in the TMR at 4\% strain, as shown in Fig.~4c.

\section{Conclusions}
In conclusion, results of our first-principles calculations show that both  SP and TMR of the CCA/NNO/CCA MTJ present asymmetrical responses to positive and negative interfacial strains. While a high SP is maintained at the interface and the TMR ratio fluctuates slightly under compressive strain, a positive strain of 4\% leads to a drop of TMR by $\sim$2 orders of magnitude due to electron tunneling through the interface states. Our study shows that a strain in the MTJ can result in significant changes in transport properties of the Heusler/semiconductor junction due to strain induced changes in coupling of electronic states across the interface. This study provides a direct evaluation of the effects of interfacial strain developed in the films on spin injection of Heusler alloy-based spintronics devices.

This work is supported by the Singapore Agency for Science, Technology and Research (A*STAR)
grant (SERC Grant No. 0921560121 and No. 0721330044).


\begin{thebibliography}{}
\bibitem{Conductance_Details}
TMR is calculated using Quantumwise package. Single-$\zeta$ polarized basis and GGA-PBE pseudopotential are adopted in transport calculations. The zero-bias conductance is evaluated by integrating the electron transmission for states at the Fermi level over the 2D Brillouin zone using a uniform $200\times200$ $k_{\sslash}$ mesh.
\end{thebibliography}
\end{document}